\documentclass [12pt,amsmath, amssymb, aps, bbm, nofootinbib, noindent, notitlepage] {article}
 \usepackage{lipsum}
\usepackage[margin=1.75cm]{geometry}
\usepackage{amssymb}
\usepackage{graphicx}
\usepackage{amsmath}
\usepackage[hidelinks]{hyperref}
\usepackage{array}
\usepackage{verbatim} 
\usepackage{multirow}
\usepackage{manfnt}
\usepackage{mathrsfs}
\usepackage{hyperref}
\usepackage{float}
\usepackage{caption}
\usepackage{braket}
\usepackage[sort, numbers]{natbib}
\usepackage{authblk}
 \usepackage{lipsum}

\usepackage{csquotes}

\numberwithin{equation}{section}

\usepackage{tocloft}

\usepackage{tocloft}

\def\be{\begin{equation}}
\def\ee{\end{equation}}

\usepackage{authblk}
\usepackage{color}

\title {TGFT condensate cosmology as an example of spacetime emergence in quantum gravity}
\author {Daniele Oriti}
	\affil{Arnold Sommerfeld Center for Theoretical Physics, \\ Ludwig-Maximilians-Universit\"at München \\ Theresienstrasse 37, 80333 M\"unchen, Germany \\ daniele.oriti@physik.lmu.de}

 \date{}

\begin{document}

\maketitle

\begin{abstract}
We summarize the main ideas behind TGFT condensate cosmology and sketch the technical steps that bring from the fundamental theory to the effective cosmological dynamics. This framework is presented as an explicit illustration of (and possibly a general template for) the emergence of spacetime from non-spatiotemporal quantum entities in quantum gravity, and the many aspects involved in it.
\end{abstract}

\section{Introduction}
The general issue that this contribution is concerned with is the nature of space and time in quantum gravity and, more specifically, the sense in which they can be said to be emergent notions in a non-spatiotemporal fundamental theory. 

This is a possibility that has been raised often and in many different contexts in the (recent) past, it is increasingly discussed in the philosophy of physics literature \cite{Huggett2021-HUGOON-3,Crowther2020-CROABS-4,Baron2019-BARTCC-15,LeBihan2019-LEBHWL,Huggett2018-HUGTTE,LeBihan2018-LEBSEI-2,Crowther2014-CROAOO} and partially realized (to a different degree) in several quantum gravity formalisms. 

First of all, we summarize what we mean by emergence in general and the different types of emergence that can be recognized in physics, and then articulate our perspective on the different \lq levels\rq or senses in which we can specifically speak of emergence of spacetime in quantum gravity. 
This serves as a necessary conceptual background for the actual focus of this work. We illustrate how space and time can be shown to emerge in the quantum gravity formalism of tensorial group field theories (TGFTs), specifically in the class of models with richer \lq quantum geometric\rq ~content called group field theories, at least in a well-identified given approximation and for the simple case of cosmological (i.e. homogeneous and isotropic) spacetimes.  


We emphasize throughout the exposition that this particular set of results may represent a concrete example of a more general template that is applicable to other quantum gravity formalisms, thus having a much more general interest. In fact, some of the technical results we illustrate are already applicable to other quantum gravity formalisms, thanks to the fact that they share several ingredients with TGFTs. Moreover, similar results, about one aspect or the other of the spacetime emergence that we illustrate in TGFTs, have been also obtained in other quantum gravity contexts.

\section{Emergence of spacetime in quantum gravity}
\subsection{Emergence and its many kinds}
We use a notion of emergence that is simple and general enough to accommodate all known examples of emergent phenomena in physics  \cite{Butterfield2011-BUTLID, Butterfield2011-BUTERA}: a physical behaviour or phenomenon is understood as {\it emergent} if it is sufficiently novel and robust with respect to some comparison class, usually associated to the class of behaviours and phenomena it emerges from. This definition can be refined in specific contexts if needed, but it suffices for our purposes. 

Two important points need to be noted. First, very often some sort of approximation or limit (in the mathematical model(s) describing the phenomena under consideration) is needed to realize such emergence \cite{pittphilsci8554, Batterman2011-BATESA, Batterman2004-BATCPA, Castellani2000-CASREA}. Second, in this definition, emergence is not  incompatible with reduction (understood, within the same mathematical model(s), as deduction); Indeed, reduction is usually needed to specify the comparison class entering the definition, so that one could even argue that  emergence is the inverse process of reduction, at the epistemological level, i,e, that we recognize some phenomenon as {\it emergent} from another exactly {\it because} it has been shown to be deducible from the other.

This contribution deals with physics, so we phrased the notion of emergence in physical terms. Associated with this physical notion, however, come two different types of emergence: intertheoretic and ontological emergence. We speak of a set of physical phenomena as emergent from another, if the theoretical description of the former can be reduced to the one of  the latter. This intertheoretic (or epistemic) emergence amounts in fact to a relation between mathematical and conceptual models of the world, from which we imply a relation between natural phenomena described by those models. Ontological emergence is instead a relation between entities themselves, thus it is both a physical and metaphysical statement. However, the philosophical perspective on which we base our discussion accepts the following points. First. all the physical entities that we assign ontological status to are to a large extent defined by the theoretical frameworks/models we use to describe them, in the straightforward sense that the concepts we use to characterize them and the properties we assign to them are those taken from such theoretical models, and not identified independently from them. I take this to follow from some basic naturalistic attitude toward metaphysics, according to which metaphysics follows from (or should be at least strongly constrained by) physics, but it is also more or less the (often naive) attitude of working physicists.\footnote{Notice that this does not imply in any way that "all the entities playing a role in physical theories should be assigned ontological status", but only the weaker (almost converse) statement that all entities to which we assign ontological status play a role in physical theories that, in fact, define them.} Second, given the above, ontological emergence follows from intertheoretic emergence, once we decide to add an ontological commitment about (some of) the entities featuring the two theories related by the emergence relations. The ontological commitment is not a trivial step by any means, and should be argued for convincingly on a case by case basis. However, once this is agreed, there is not much left to argue about the existence of some ontological emergence. Obviously, non-trivial metaphysical consequences can then follow from the existence of ontological emergence, and this adds to the non-triviality of any ontological commitment. In this contribution, we focus on the current and future theoretical models of space and time, and thus on intertheoretic emergence of spacetime, and on the more physical rather than metaphysical aspects. Thus, we limit ourselves to emphasize at which level the intertheoretic relations make room for ontological emergence and what sort of emergence this could be, once the ontological commitment is agreed upon. We choose not to discuss the very many additional subtleties concerning the conditions for an ontological commitment to be justified and, more generally, concerning the complex relations between theoretical models and concepts, models and reality, including the definition of the latter, the normative vs representational character of natural laws and how these are captured by theoretical models, and many more. 

Note that any intertheoretic emergence, and thus any ontological emergence in our (\lq naturalistic\rq) perspective, is in itself not associated to any physical process happening in time and leading from the fundamental to the emergent description (and/or entity). It corresponds rather to a change in perspective, accompanied by a change in relevant concepts (and entities inferred in correspondence to such concepts). It is, then, a {\it synchronic} emergence, not corresponding to a temporal process, a priori.\footnote{A much better term would probably be "a-chronic", though, since "synchronic presupposes the existence of some time.} This is the case also for the peculiar example of emergent behaviour often associated to phase transitions, whose peculiarity lies in the fact that they are {\it symmetric} relations between theoretical descriptions (and corresponding entities) in which none can be said to be more fundamental than the other (while both could be emergent from some more fundamental \lq microscopic \rq theory). 

While this is not part of the definition, however, it is not excluded either, and many examples of physical processes can be found, that instantiate emergent behaviour, then characterized in intertheoretic terms. In this sense, the same emergence that was understood in synchronic terms at the intertheoretic level can then be understood in {\it diachronic} terms as something happening {\it in time}.\footnote{Phase transitions are a case at hand, but even the transition from molecular dynamics to hydrodynamics (in itself not a physical process) could be associated to specific dynamical processes of molecules whose collective behaviour then requires the switch to an hydrodynamic description, so that one could say that the emergence of fluids and fluid dynamics takes place {\it in time} along these dynamical processes.} In the case of the emergence of time itself, by definition a temporal characterization is ruled out, which makes the understanding it even more challenging, and one can at best imagine a partial temporal characterization \lq up to the time when time disappears\rq \cite{Oriti:2021zju}. We will come back to this point when discussing our specific example of spacetime emergence.

\subsection{Spacetime emergence and its many levels}

Let us now summarize the \lq levels of emergence \rq that can be envisioned for spacetime in quantum gravity\footnote{Specific quantum gravity formalisms of course will exemplify the framework, which is meant to apply in full generality, in different ways and with their own peculiarities.}. A more detailed outline can be found in \cite{Oriti:2018dsg} and, with a focus on the emergence of time, in \cite{Oriti:2021zju}.

\

General relativity is a dynamical theory of continuum (in fact, smooth) fields defined on a (differentiable) manifold (i.e. a set of points with appropriate regularity properties). Thus, at first the underlying ontology seems to be given by these elements: fields (including the metric) and manifold. However, diffeomorphism invariance of the theory implies that values of fields at different points in the manifold can be physically equivalent and thus the manifold and its points do not really carry any physical meaning in themselves and they are thus not part of the ontology of the world, except as providing global (topological) conditions on the fields (since the set of allowed field configurations depends on the manifold topology).\footnote{This conclusion, although the most reasonable and well-supported one in our opinion, is not uncontroversial and the debate about the nature of spacetime in classical GR, but also about the role of the manifold, for better or worse, still goes on \cite{Pooley2013-POOSAR, Hoefer1998-HOEAVR}.} Moreover, the theory is background independent in the sense that all fields appearing in it are dynamical entities \cite{Giulini:2006yg, Gaul:1999ys}, subject to \lq equations of motion\rq~ constraining their allowed values and mutual relations. And generic solutions possess no feature that can be used to single out a preferred direction of time or space, that can only be associated to geometries with distinctive isometries or to special boundary conditions (or both, like AdS spacetimes). 

In this sense (absence of preferred time or space) one could claim that there is in fact no space or time in GR: there are only fields and their relations. 
In fact, just like physical (i.e. diffeo-invariant) observables can be constructed as relations between dynamical fields, time and space can be defined by choosing appropriate (matter) fields (components) as rods and clocks and computing spatial and temporal quantities using them and the metric field (which enters all such quantities, and in this sense encodes the physical, dynamical spacetime in GR). This is the {\it relational strategy} \cite{Rovelli:1990ph,Marolf:1994nz,Rovelli:2001bz, Dittrich:2005kc} to the definition of space and time in GR, and its application can be said to implement a first kind of {\it emergence} of space and time within a theory that does not select one a priori. We could call it a {\it level $-1$} of spacetime emergence. Indeed, generic fields will not behave like a perfect rod or clock and the notions of space and time that they will concur to define will not match the usual notions (basically corresponding to the Newtonian ones of our common sense), which is only the case in some special approximation and for special kinds of fields.

Special kinds of fields are also introduced and used in classical GR in order to allow for the {\it deparametrization} of the theory, i.e. its formulation in an entirely diffeo-invariant language and with a clearly identified notion of space and time (that's the main benefit of this procedure), in the (preferred) frame defined by such material rods and clock  (that's the main shortcoming). One can then proceed to quantization for example by standard canonical quantization methods, assuming that the frame degrees of freedom are not quantized, and obtain a rather standard quantum field theory for the other fields, including the metric \cite{Husain:2011tk, Thiemann:2004wk,Giesel:2012rb,Giesel:2017roz}. Provided one can complete the quantization procedure (at present something that is only achieved in a rather formal sense), we end up with a theory of quantum gravity which could be valid as long as the quantum dynamical nature of the matter fields chosen as preferred reference frame can be neglected (again, we see that we have the usual notions of time and space only as the result of some idealization or approximation).

\

Because the metric is quantised (and dynamical), however, space and time do not remain what they were even in presence of a preferred reference frame. 
In fact, in a quantum theory of gravity obtained from the quantization of the metric field \cite{Kiefer:2013jqa,Kiefer:2012cdl, Thiemann:2007pyv}, they {\it disappear} in a more radical sense in which they had already disappeared in classical GR. Accordingly, the emergence of space and time from quantum gravity presents much more radical challenges that the one in classical GR. It does not imply any ontological emergence, since the fundamental entities remain the same, i.e. dynamical fields, and relational construction remain essential to identify physical notions of time and space. The quantum description of fields, including the metric field, however, implies that these notions (and all geometric ones: distances, curvature, volumes, causal relations) will be subject to uncertainty relations, irreducible quantum fluctuations, some form of contextuality, discreteness of observable values, and, in the case of composite systems, entanglement. The challenges to our common sense notions of realism, separability, and locality are formidable, even more than in standard quantum mechanics. 

The relational strategy for the definition of space and time will be affected by the quantum properties of our relational frames, in particular our clocks \cite{Hoehn:2019fsy,Castro-Ruiz:2019nnl,Giacomini:2021gei}. We must abandon any value-definiteness of spatiotemporal quantities and possibly any continuous notion of space and time, if spatiotemporal observables end up having discrete values \cite{Rovelli:1994ge}. We have to learn to deal with quantum reference frames and indefinite causal structures or temporal order \cite{Hardy:2005fq, Castro-Ruiz:2019nnl,Giacomini:2021gei} (see also the chapter by Lam, Letertre, and Mariani in this volume), and we are forced to abandon unitary time evolution as a key aspect of quantum dynamics. All these aspects are of truly foundational nature as well as bearer of important physical consequences, but have not been studied as much as they deserve in a full quantum gravity setting, even in quantum gravity contexts like the TGFT one we will discuss in the following where quantum reference frames are indeed used to extract a relational dynamics from non-spatiotemporal entities.
Starting from such quantum realm, the emergence of space and time as we know them from GR requires a number of approximations and restrictions, which together define the semiclassical limit of quantum gravity. This includes the use of special semiclassical states, the focus on a subset of observables, and more. 

The very presence of this additional step justifies speaking of a new level of emergence, level 0, common to any quantum theory of gravity and spacetime obtained by quantization of the classical one. Once more, we are dealing with an intertheoretic, synchronic emergence. Whether any physical process (a \lq classicalization process\rq of spacetime and geometry) can be put in correspondence with it can only be determined within some specific quantum gravity formalism. Even if it can, speaking of the emergence of spacetime (time, in particular) as if it was a temporal process remains impossible (outside special approximations) \cite{Oriti:2021zju} as anticipated above.

\

A new level of spacetime emergence is found in quantum gravity formalisms in which the fundamental theory is not understood as the straightforward quantization of the classical gravitational theory, because it is instead based on a new set of basic entities that do not correspond simply to the quantum counterpart of the continuum fields of GR \cite{Oriti:2018dsg}. This is an intertheoretic spacetime emergence of {\it level 1}, which by definition corresponds to an ontological emergence as well, since the spatiotemporal fields of GR (i.e. those, including the metric, whose relations define space and time as we know them) are replaced by non-spatiotemporal entities. The precise nature of such new entities varies in different quantum gravity formalisms, as does the precise degree by which they differ from continuum fields. We will deal in some detail with one specific example in the following. 

The step away from the very ontology of fields brings us new conceptual and technical challenges with respect to the already challenging disappearance of spacetime forced upon us by quantum spatiotemporal fields in any quantized version of GR. 
There is a new non-spatiotemporal ontology to make sense of, first of all; and the need to make sense of the ontological status of spatiotemporal fields, of spacetime itself, now that they are deprived of fundamental status. The precise dependence relation between fundamental entities and emergent spatiotemporal fields needs to be clarified, the best options being in the sense of functionalism or grounding, indeed apt to be applied also in the analogous context of fluid dynamics emerging from atomic or molecular physics. At the technical as well as physical level, on the other hand, the big challenge is to control the collective behaviour of the new quantum entities replacing fields, since the latter, including the metric field and any associated notion of space and time, should arise from such collective behaviour, as coarse grained, approximate notions. 

This is often referred to as the problem of the \lq continuum limit\rq in quantum gravity, since the new fundamental entities are often some discrete (and quantum) counterpart of continuum fields, and it is usually tackled by renormalization group methods and other tools form statistical and quantum many-body theory \cite{Bahr:2017klw,Dittrich:2014ala,Eichhorn:2018phj,Carrozza:2016vsq,Finocchiaro:2020fhl}. It is important to stress that such continuum limit is conceptually as well as technically distinct from the classical one, which would be the only one responsible for the emergence of spacetime from quantum gravity at level 0. Indeed, working at this level 1, we could expect the continuum limit to bring us to level 0, with the emergence of spacetime as we know it from GR still requiring the additional steps we outlined above.   

\

The situation becomes more complicated still, once we realize that the continuum limit of a quantum many-body theory of non-spatiotemporal entities is not unique, in general. 
We should rather expect it to give rise to several continuum phases of the same fundamental quantum gravity system. Since we have to match (at least approximately) the well-tested general relativistic description, at least one of these continuum phases should be geometric or spatiotemporal. That is, it should allow for an effective reconstruction of continuum spacetime and geometry, after further (classical) approximation (and a relational rewriting of the theory). But other will be non-geometric and non-spatiotemporal, not allowing such reconstruction, even approximately. 

This is a \lq level 2\rq emergence, since it clearly brings further issues to be solved. At the physical level, they concern the study of the phase diagram of the fundamental quantum gravity theory and the identification of the one (or more) phase(s) in which we reach a situation corresponding to a level 0 emergence of spacetime, i.e. a situation like in quantum GR (if not directly a level $-1$ emergence, that of classical GR, if the classical approximation is somehow part of the continuum limit). At the conceptual level, new difficulties arise because the same kind of non-spatiotemporal entities that require a new ontology (one that does not make use of spatiotemporal notions) are recognised to be \lq even less spatiotemporal\rq now, since they may not give rise to spacetime at all, even in a continuum approximation \cite{Oriti:2018dsg}. 

\

This is another point where the ontological commitment one is willing to adopt concerning the new non-spatiotemporal structures appearing in quantum gravity formalisms changes drastically the conceptual context in which the same formalisms are understood.

The non-spatiotemporal entities adopted as basic structures in several quantum gravity may be considered purely mathematical tools, instrumental for defining a quantum theory of the gravitational field alongside other matter fields, assumed to be the only truly physical entities. This does not change at all the technical difficulties to be solved, with the continuum limit, the coarse graining and renormalization techniques, the non-geometric phases, etc, but it implies that physics is found only in level 0 of spacetime emergence, and that only at the same level we should be concerned with the associated conceptual difficulties. 

If one takes seriously the same entities as (potentially) physical ones, on the other hand, taking seriously also the probably rich phase diagram of the fundamental theory becomes necessary. One is also led to consider the possible physical nature of the phase transitions separating geometric and non-geometric phases. What is the physical interpretation and what are the physical signatures of a {\it geometrogenesis} phase transition \cite{Konopka:2006hu,Oriti:2007qd,Mandrysz:2018sle, Smolin:2003qf} \cite{Delcamp:2016dqo, Kegeles:2017ems, Bahr:2015bra, Koslowski:2011vn, Surya:2011du, Pithis:2020kio,Ambjorn:2016mnn}\cite{Percacci:1990wy}, leading from a non-geometric phase to a geometric one or viceversa? What are the relevant geometric order parameters? For example, can the cosmological big bang singularity of classical GR be in fact replaced by such geometrogenesis phase transition in quantum gravity? Can we locate {\it in time} (e.g. in our cosmological past) this emergence of time itself? And what is the physical interpretation of the other quantum gravity phase transitions (and of the other phases)? 

New issues appear, at both physical and philosophical level, deepening further the challenge of making sense of spacetime emergence. Because of them, we identify this situation as a further level, i.e. \lq level 3\rq ~of spacetime emergence in quantum gravity.  

In the end, assigning ontological status to the new non-spatiotemporal entities suggested by quantum gravity approaches calls for the development of an ontology that is not spacetime-based, which is a philosophical work yet to be tackled, and of the outmost importance.

\

A schematic representation of the multi-level scenario we outlined is given below.

\begin{figure}[h]
\includegraphics[width=20cm]{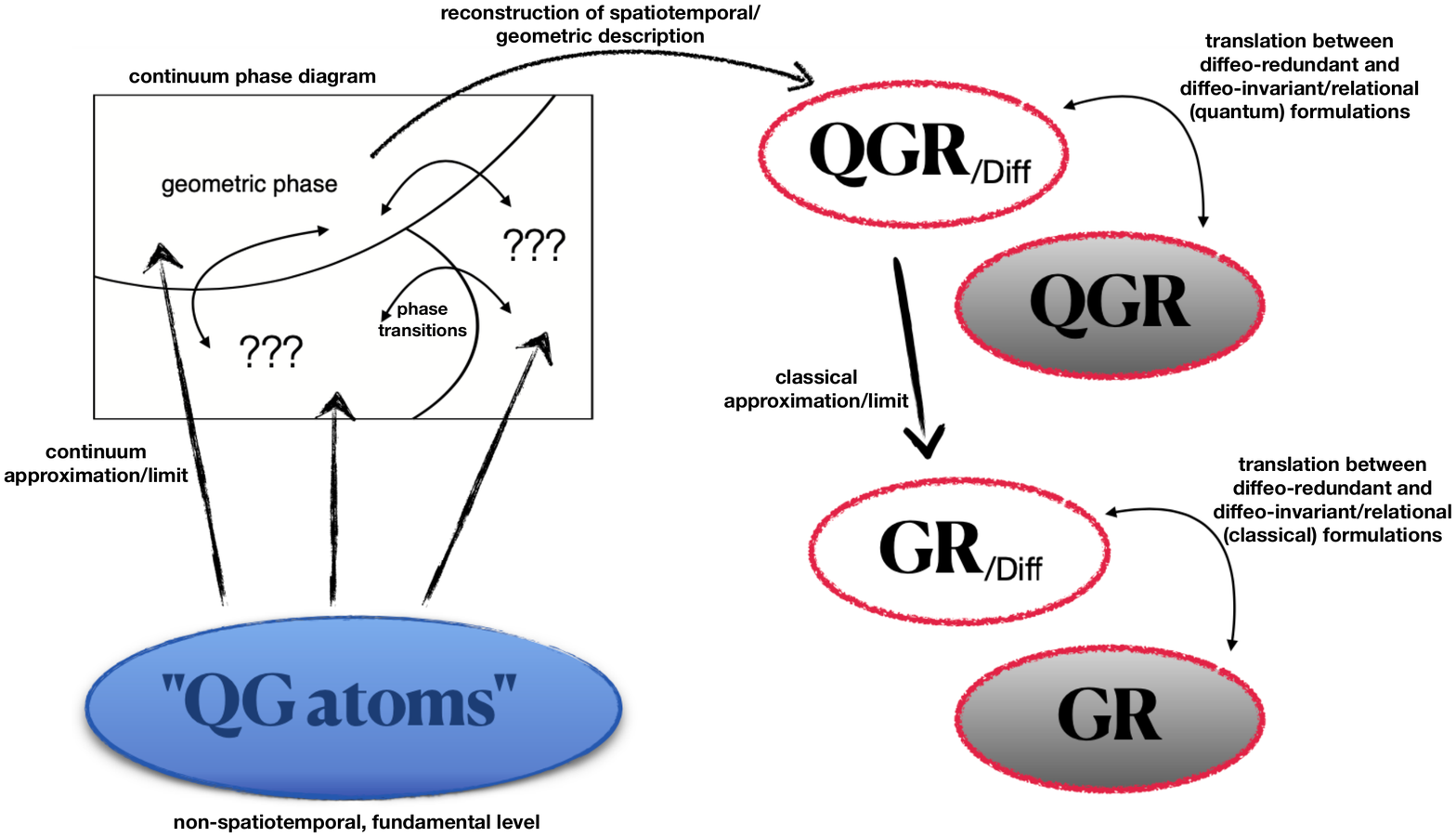}
\caption{A schematic representation of the multi-level scenario for spacetime emergence.}
\label{fig:emergence}
\end{figure}

\

It should be clear that both physical and philosophical challenges raised by this quantum gravity scenario are of the deepest nature. We will not discuss them here. Instead, we proceed to present a concrete example of a quantum gravity formalism in which all these levels of spacetime emergence have been realized, tentatively and partially, in recent work.

\section{The TGFT atoms of space}
The example of spacetime emergence we discuss is realized in the context of tensorial group field theories \cite{Oriti:2011jm, Gurau:2012hl,Krajewski:2012aw,Rivasseau:2016wvy, Oriti:2014uga}, and more specifically of the most \lq quantum geometric\rq subclass of such models, usually called group field theories. We also restrict attention to 4-dimensional Lorentzian models, i.e. the most directly relevant for physics.

\

The fundamental entities of these quantum geometric models are quantized tetrahedra whose discrete geometry is encoded in algebraic data. Specifically, the Hilbert space associated to an individual tetrahedron is obtained starting from $\mathcal{H} = L^2\left( G^4; d\mu_{Haar}\right)$ and imposing appropriate additional \lq geometricity \rq restrictions on quantum states. The Lie group $G$ is the Lorentz group $SO(3,1)$ (or its double cover $SL(2,\mathbb{C})$) or its rotation subgroup $SU(2)$, $d\mu_{Haar}$ is the corresponding Haar measure. The geometricity conditions can also be imposed at the dynamical level. 

In this representation of the quantum tetrahedra, quantum states are then functions of group elements, each associated to one of the triangles on the boundary of each tetrahedron. One can also use different equivalent representations of the same Hilbert space, for example in terms of (unitary, irreducible) group representations associated as well to the boundary triangles, so that one can represent the same tetrahedron equivalently as a spin network vertex, i.e. a vertex with four outgoing open links, each labeled by a group representation. This representation of the tetrahedral quantum geometry can be obtained by straightforward quantization of a phase space description of the (intrinsic and extrinsic) classical geometry of the same tetrahedra in terms of same kind of group-theoretic data \cite{Baez:1999tk, Bianchi:2010gc, Dittrich:2008ar, Baratin:2011hp, Rovelli:2010km, Perez:2012wv, Finocchiaro:2020xwr}. 

Building on this single-tetrahedron Hilbert space, one can then define a Fock space for the quantum states  of arbitrary numbers of tetrahedra: $\mathcal{F}\left( \mathcal{H}\right) = \bigoplus_{V=0}^{\infty}sym\left\{ \mathcal{H}^{(1)}\otimes\mathcal{H}^{(2)}\otimes \cdots\otimes\mathcal{H}^{(V)}\right\}$. Field operators creating/annhiliating the quanta of this Fock space, i.e. quantum tetrahedra, can then be straightforwardly introduced, to realize a field theory formulation of pre-geometric \lq atoms of space\rq.

This Fock space contains quantum states associated to extended simplicial complexes formed by gluing tetrahedra to one another across shared boundary triangles (or equivalently, associated to extended 4-valent graphs obtained by gluing open vertices to one another across their open links). The elementary gluings are obtained as maximal entanglement between the quantum degrees of freedom associated to the same triangle in the two tetrahedra to be glued (and correspondingly for the dual graphs) and the overall connectivity of such discrete structures is encoded as entanglement patterns associated to the quantum states in the TGFT Fock space \cite{Colafranceschi:2020ern}. 
\begin{figure}[h]
\begin{minipage}[t]{8.5cm}
		\centering
		\includegraphics[width=4cm]{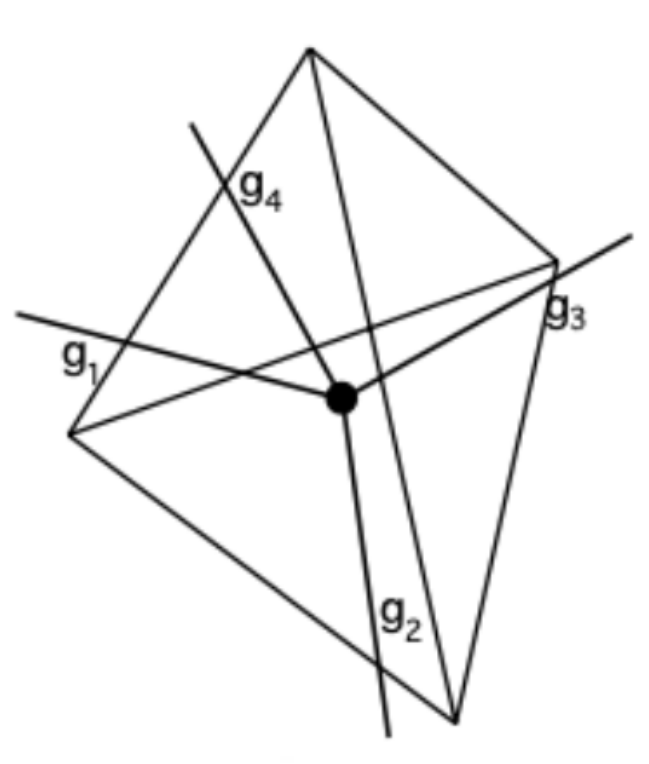}\caption{A tetrahedron with its dual graph representation, labelled by group elements.}
		\label{tetrehedron}
	\end{minipage}
\hspace{0.5cm}
\begin{minipage}[t]{8.5cm}
		\centering
		\includegraphics[width=6cm]{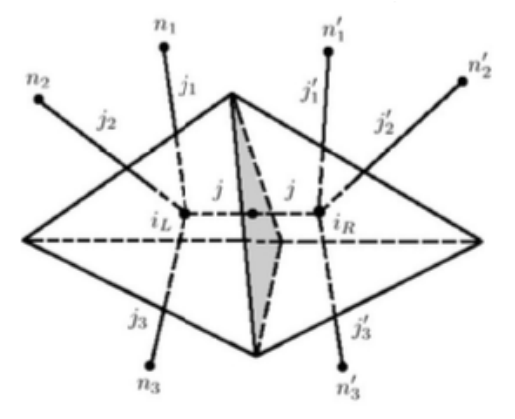}
		\caption{Two tetrahedra with the dual graph representation, glued along a shared triangle, labelled by group irreps.}
		\label{gluedtetrahedra}
	\end{minipage}
\hspace{0.5cm}	
\end{figure}

In terms of group irreps for $G=SU(2)$, the resulting connected states match the spin network states of canonical loop quantum gravity \cite{Sahlmann:2010zf} and the Hilbert space associated to them, for a given graph, coincide. When considering quantum states associated to arbitrary graphs and their scalar product, i.e. the full Hilbert space of the theory, they differ, with the one in canonical LQG encoding more conditions motivated by continuum geometry, absent in the Fock space of TGFT \cite{Oriti:2013aqa,Oriti:2014uga}. In this sense, quantum geometric TGFTs are a 2nd quantized version of LQG, with a stronger discrete flavour.   

\

So there is a clear connection between TGFT states and discrete (piecewise-flat) geometries, at least for some of them and in a semiclassical approximation, and this discrete geometric intuition guides model building and the analysis of their quantum dynamics. It is also clear, however, that they are not at all fully spatiotemporal structures. Generic TGFT states correspond to arbitrary numbers of (partially) disconnected quantum simplices, themselves not behaving like classical geometric ones, even in the limited sense of piecewise-flat geometry. Moreover, even the very special ones with a nice simplicial geometric interpretation (thus corresponding to quantum piecewise-flat geometries) will remain distant from (quantized) continuum geometries. The list of \lq geometric or spatiotemporal pathologies\rq ˜that generic quantum states for arbitrary collections of quantum TGFT building blocks can possess is long and it marks a huge gap with respect to the usual description of spacetime in terms of fields (including a metric field), a gap that is much wider than the already relevant gap between classical piecewise-flat geometry for extended simplicial complexes and the smooth geometry of classical GR. This justifies referring to this level of description as non-geometric and not spatiotemporal.  Of course, we have to go beyond the use of the word \lq geometric\rq ˜in both contexts, realizing that the corresponding sense in which we have \lq geometry\rq ˜is very different in the two cases, and that our usual notions of space and time only apply to one of the two, really.

\

The same considerations apply at the dynamical level. 

A partition function for the TGFT \lq atoms\rq, i.e. the quantized tetrahedra, and for the field over the group manifold of which they represent elementary quanta, will in general take the form:

\begin{equation}
Z = \int \mathcal{D}\varphi\mathcal{D}\varphi^*\, e^{- \, S_\lambda(\varphi,\varphi^*)} \qquad ,
\end{equation}
for an action $S_\lambda(\varphi,\varphi^*) = K + U + U*$ function of the field (and complex conjugate) and some coupling constant(s) weighting an interaction term $U$ given by a polynomial in the fields and encoding a pairing of their (group) arguments which is, in general, non-local (in the sense that they are not simply identified at interactions, like positions in usual local QFT) \cite{Gurau:2012hl,Oriti:2011jm,Krajewski:2012aw,Rivasseau:2016wvy, Oriti:2014uga}. $K$ is instead a quadratic (in the fields) kinetic term.

It can be defined in quantum statistical mechanics setting in terms of entropy maximization for given thermodynamic constraints on the fundamental TGFT quanta \cite{Chirco:2018fns, Kotecha:2019vvn}. These constraints can be motivated by discrete gravity considerations or as corresponding to a dynamical (Hamiltonian) constraint of a canonical (loop) quantum gravity framework. 

The best established connection between the TGFT quantum dynamics (for quantum geometric models) and that of discrete quantum gravity and loop quantum gravity is to view the TGFT partition function as the generating function for a discrete gravity path integral, incorporating also a sum over discrete topologies. In formulae:

\begin{equation}
Z = \int \mathcal{D}\varphi\mathcal{D}\varphi^*\, e^{- \, S_\lambda(\varphi,\varphi^*)} = \sum_{\Gamma} \frac{\lambda^{N_\Gamma}}{sym(\Gamma)}\,\mathcal{A}_\Gamma \qquad .
\label{eq:Zpert}
\end{equation}

Indeed, for appropriately quantum geometric models as the ones we focus on, the TGFT partition function can be expanded perturbatively with respect to its coupling constants (weighting the interaction terms in the TGFT action), the generic result being that: a) the Feynman diagrams $\Gamma$ so generated as dual to cellular (simplicial) complexes of arbitrary topology; b) the (model dependent) Feynman amplitudes $\mathcal{A}_\Gamma$ are given by lattice gravity path integrals for a (Plebanski-like) formulation of discrete gravity in terms of group-theoretic data, on the lattice dual to the Feynman diagram $\Gamma$ \cite{Baratin:2011hp, Finocchiaro:2018hks}; c) the same Feynman amplitudes $\mathcal{A}_\Gamma$, when expanded in irreps of the group $G$, take the form of spin foam models \cite{Reisenberger:2000zc,Perez:2012wv,Finocchiaro:2018hks}, which in turn can be understood as a covariant formulation of the quantum dynamics of the spin network states of canonical loop quantum gravity. 

The first point is common to all TGFT models, and a large number of results in simpler models (in particular, random tensor models) clarified combinatorial, topological and non-perturbative aspects of the sum over complexes they generate \cite{Gurau:2012hl, Rivasseau:2016wvy}. These simpler TGFT models also share the second point, but in general their Feynman amplitudes can be understood as discrete gravity path integrals only in the purely combinatorial sense of dynamical triangulations \cite{Loll:2019rdj}. The additional group-theoretic data of quantum geometric TGFT models enrich this discrete gravity description with dynamical geometric information, like in quantum Regge calculus \cite{Hamber:2009mt} and spin foam models, and are necessary for the third point to be realised. 

The considerations we put forward at the kinematical level, concluding that the basic TGFT entities should be understood as pre-geometric, non-spatiotemporal entities, apply also at this dynamical level. The TGFT partition function can be seen as the generating functional for a gravitational path integral provided one can give non-perturbative and continuum meaning to the discrete structures appearing in its initial definition. 

These are still far from a theory of continuum fields and spacetime and in fact generic discrete configurations appearing in the perturbative expansion would not even correspond to well-behaved simplicial geometries. Still, this connection to discrete gravity path integrals and to spin foam models is very important for TGFTs. It clarifies the interpretation and meaning of the TGFT quantum dynamics, it provides a key guide for model building, and it illustrates the strong ties to other quantum gravity approaches which in turn allow importing (and exporting) techniques and results. 

\

The guideline provided by the connection to discrete gravity path integrals is crucial also for extending TGFT models to matter coupling. In particular, we are interested in adding degrees of freedom that can be interpreted as discretized scalar matter,\footnote{The use of scalar fields can be seen as a simplifying choice to actually mimic more realistic matter content (e.g. an actual set of rods and clock, if one wants to use them as relational frames).} just like the group-theoretic variables can be interpreted as discrete geometric data. Thus, the model building strategy is to define a TGFT field and action in such a way that, in perturbative expansion, one obtains Feynman amplitudes with the form of discrete path integrals for gravity coupled to scalar fields, on the lattice dual to the Feynman diagram.
Beside the obvious physical relevance of matter coupling, this extension of TGFT models is needed in order to apply the relational strategy for the reconstruction of space and time observables from the full theory. 
 
 Focusing on a single real scalar field (to be later used as a relational clock), one works with a TGFT field $\varphi\left( g_I , \chi\right)$ defined on an extended domain $G^4 \times \mathbb{R}$ and with an action of the general form \cite{Oriti:2016qtz,Li:2017uao}:
\begin{eqnarray} \label{GFTaction}
S_\lambda(\varphi,\varphi^*) \,&=& \,K \,+ \,U \,+ \,U^* \\
K &=& \int dg_I dh_I \int d\chi d\chi' \varphi^*(g_I;\chi) K\left( g_I,h_I; (\chi - \chi')^2\right) \varphi(h_I;\chi') \nonumber \\
U &=& \int [dg] d\chi \, \varphi(g;\chi) ... \varphi(g';\chi)\,\mathcal{V}\left( \{ g\} ; \chi \right) \nonumber \qquad ,
\end{eqnarray}
with a kinetic term involving some (local) dependence on quantum geometric data and that can be expanded in (infinite) powers of second derivatives of the TGFT field with respect to the scalar field variable $\chi$, and an interaction term that is local in the scalar field data, while remaining non-local in the quantum geometric ones.

Notice that the scalar field variable (to be later used as a clock to define relational observables) enters the TGFT action in a way compatible with its use as a standard time coordinate in the analysis of the field theory itself (provided one can deal satisfactorily with the higher derivatives). This led to develop a TGFT counterpart of the \lq deparametrization\rq strategy of classical GR, and a straightforward canonical quantization of TGFT models (and consequent cosmological applications) \cite{Wilson-Ewing:2018mrp, Gielen:2019kae, Gielen:2021dlk}.

\

TGFT models propose therefore a formulation of quantum gravity in terms of non-spatiotemporal \lq atoms of space\rq, an example of the \lq QG atoms\rq in Fig~\ref{fig:emergence}, from which continuum spacetime and geometry should then be shown to emerge.

Let us now turn to how this emergence is realized in this formalism.

\section{Continuum limit in TGFT models}
In a quantum field theory formalism, albeit peculiar like the TGFT one, the full quantum theory incorporating all dynamical degrees of freedom (and arbitrary numbers of the fundamental quanta) and thus also the \lq continuum limit\rq , corresponds to the evaluation of the full partition function $Z$ (or free energy $F = \ln Z$) or, equivalently, the full quantum effective action $\Gamma[\phi] = sup_J\left( J \cdot \phi - F(J) \right)$ for source $J$ and mean field $\phi = \langle \varphi\rangle $. 

This is also in line with the discrete gravity interpretation of the TGFT quanta and perturbative amplitudes. The evaluation of the TGFT partition function or effective action amounts to resumming the full perturbation series, thus the sum over triangulations weighted by a discrete gravity path integral in ~\ref{eq:Zpert}, including infinitely refined lattices. This is also how one would define the continuum gravitational path integral starting from the discrete one in simplicial quantum gravity approaches, as well as in spin foam models. The caveat, with respect to the discrete quantum gravity intuition, is that the TGFT picture suggest that discrete gravity and spin foam models may capture only part of the QG story, and thus to look beyond the perturbative regime, exploring the corresponding TGFT models in the full non-perturbative regime, i.e. for large values of the couplings too. 
In physical terms, this means being able to control the full collective quantum dynamics of the QG atoms, looking for regimes in which the discrete picture can (and should) be replaced by one in terms of continuum spatiotemporal fields. 

\

Much recent work has been devoted to this issue, for different TGFT models and from different perspectives. 
A main tool is the renormalization group. Besides a number of constructive techniques proving that, at least for simpler TGFT models, a rigorous non-perturbative definition can be achieved, most results have been obtained employing functional renormalization group methods. Fully quantum geometric models have proven so far too involved to be treated at such non-perturbative level, and we are only recently gaining some understanding fo the scaling properties and basic divergence structure, from a perturbative point of view. On the other hand, simpler TGFT models have been analysed in detail, for different choice of domain, rank, features of the dynamics, thanks also to the detailed knowledge about the underlying combinatorial structures gained in the even simpler context of random tensor models. We have many examples of perturbatively renormalizable TGFTs and a good understanding of their RG flow, at least for suitable truncations in theory space. In particular, we are gaining important knowledge about the conditions leading to interesting critical behaviour and phase transitions \cite{Pithis:2020kio}. Reviews about this body of work can be found in \cite{Carrozza:2016vsq,Baloitcha:2020idd, Finocchiaro:2020fhl}. 

\

From the point of view of spacetime emergence in fig.~\ref{fig:emergence}, we are then trying, step by step, to complete the move from QG atoms to the (mathematical) definition of the continuum phase diagram of the fundamental theory, which would amount to be able to control the (formal properties of the) full theory, i.e. the dynamics of the QG atoms at all scales and in all regimes. 

Tackling physical issues, of course, requires much more than formal mathematical control. We need to identify which continuum phase admits a rewriting of the theory in terms of spatiotemporal fields and a GR-like dynamics, at least approximately, and then do new physics starting from the fundamental theory.  

We need more than full mathematical control, but also less.
Of course, while computing the full quantum effective action would be ideal, it is simply not feasible, as it is not for any quantum many-body system of the kind of complexity that we expect a full quantum gravity theory to have. We need approximation schemes that allow us to get at least a partial glimpse of the continuum phase diagram, and mathematical control over the approximation schemes, so that we are able to improve them.
And we need to extract physical insights from the partial, approximate picture we obtain in this way, without waiting for the full picture to become available. That's how physics works.

If the emergence of space and time takes place due to the collective dynamics of the QG atoms, we need approximation schemes that capture such collective dynamics, that correspond to some form of coarse-graining of the fundamental \lq atomic\rq dynamics, and that maintain visible the quantum nature of the same atoms (since the continuum limit is distinct from the classical one, and it could well be that quantum properties of the QG atoms are in fact responsible for key aspects of the spatiotemporal physics we want to reproduce). 

\

The simplest approximation of the full quantum effective action, that possesses these features, is the mean field approximation. That is, the first place to look for spacetime physics, in the perspective we are advocating, is TGFT mean field hydrodynamics. It corresponds to the saddle point evaluation of the full TGFT path integral, and to approximating the full quantum effective action with the classical TGFT action $\Gamma[\phi] = S(\phi)$. It is the natural starting point, which has then to be improved.

Notice that (despite the name), from the point of view of the QG atoms, i.e. the TGFT quanta, this still amounts to working with rather highly quantum states. Indeed, it amounts to work with coherent states $| \sigma \rangle$ of the TGFT field,\footnote{They satisfy $\widehat{\varphi}| \sigma \rangle = \sigma | \sigma \rangle$.} whose expression in terms of TGFT Fock excitations is:

\begin{equation}
| \sigma \rangle = \exp (\widehat{\sigma}) | 0 \rangle \qquad \widehat{\sigma} = \int d[g] d\chi \sigma(g_I;\chi) \widehat{\varphi}(g_I;\chi) \qquad ,
\end{equation}

where $| 0 \rangle$ is the Fock vacuum (no QG atoms at all), and the exponential operator acting on it can be expanded in power series to give an infinite superposition of states with increasing number o tetrahedra or spin network vertices (the QG atoms), all associated with the same wavefunction $\sigma$, i.e. a coherent state of infinite quantum gravity degrees of freedom. \footnote{This is exactly how the classical electromagnetic field is obtained from coherent states of photons.} 

Still, while highly quantum, they are extremely simplified states with respect to generic quantum states of tetrahedra or spin network vertices. In particular, they do not encode quantum correlations (entanglement, connectivity information) among them. They are a first step, to be then improved, being in fact the quantum gravity counterpart of the Gross-Pitaevskii approximation  in the hydrodynamics of quantum liquids. When moving to this level of description, we are then moving from the QG atoms to the full continuum description of quantum gravity, but within a specific regime of approximation, which remains quantum and focused on the collective properties of the same QG atoms, rather than their individual, pre-geometric features. Conceptually, we are solidly within a {\it level 1} emergence scenario for spacetime.

\section{TGFT condensate cosmology}
The task is now to obtain, from the fundamental QG theory, an effective dynamics that can then be understood in spatiotemporal terms, in the sense of quantum GR, i.e. an effective dynamics of geometric quantities as those constructed out of continuum fields including the metric, maybe still possessing quantum properties (as in a {\it level 0} of spacetime emergence). 
This is what TGFT condensate cosmology sets to do \cite{Gielen:2016dss, Oriti:2016acw,Pithis:2019tvp}.

\subsection{Emergence of spacetime in continuum approximation in specific phase}
The general strategy is then to look for continuum spacetime at the level of collective dynamics of the fundamental entities. 
The more specific hypothesis is that the relevant phase of the QG system is a {\it condensate phase}, guided by the intuition of the universe as a kind of condensate, indeed, of QG entities (in the same sense in which a quantum fluid is a condensate of atoms). 
The second hypothesis is that the relevant dynamical regime is the hydrodynamic one, with the relevant physical information captured by the condensate wavefunction as the main dynamical variable. 

Then, as we have just discussed, the first approximation we look at is mean field hydrodynamics (which of course amounts to the assumption that a Gaussian, weakly interacting regime is already good enough to unravel interesting spacetime, gravitational aspects).

Since we are working with an interacting quantum many-body system, albeit a peculiar one, it was immediately necessary to consider the rich continuum phase diagram that is expected in such systems, and thus the issue of which phase could be the relevant one to look at. The idea is that the definition of meaningful geometric and spatiotemporal quantities requires the use of a non-vanishing expectation value of the TGFT field operator (the condensate wavefunction or mean field), i.e. that we find ourselves in a broken, condensate phase. The symmetric or unbroken phase, with vanishing mean field as order parameter, would then correspond to a non-geometric and non-spatiotemporal phase, in which all (or most) such quantities are vanishing or somewhat pathological.

When considering also these aspects (conceptual and technical) of the story, we are working, in fact, within a {\it level 2} spacetime emergence, with the consequent more radical departure of the fundamental ontology from a spatiotemporal one, and the additional issues that we discussed above.

\

Considering now what sort of spatiotemporal physics we can extract from the TGFT hydrodynamics, for quantum geometric (GFT) models, the immediate guess is that we can only obtain cosmological dynamics, i.e. restricted to homogeneous geometries. 
There are two orders of reasons, both quite heuristic. One is that the conditions that identify the simplest hydrodynamic regime of a quantum many-body system, i.e. focusing on macroscopic, global variables only, corresponding to the maximal coarse graining of microscopic details, limit oneself to close-to-equilibrium dynamics, intuitively match the ones corresponding to the homogeneous or near-homogeneous cosmological sector of gravitational physics, at least away from the cosmological singularity. Another follows from noticing that the GFT mean field has the same domain of a wavefunction of a single GFT \lq atom\rq, i.e. an individual 3-simplex, encoding its (usually spacelike) quantum geometry, and that, by definition, it does not encode any notion of \lq local variation of geometry\rq  ~at least from the point of view of discrete gravity (as a result of maximal coarse graining).  In models extending the pure (discrete) geometric domain to matter degrees of freedom, like in ~\ref{GFTaction}, in presence of a single scalar field the best one can do is to try to use this additional degree of freedom as a relational clock to define a notion of temporal evolution, for homogeneous geometries. This is what we focus on, in the following, while for the study of inhomogeneities the relational strategy requires the introduction of more (matter) degrees of freedom in addition to the matter clock.

\

The guess turns out to be supported by the following general fact, whose validity goes beyond specific models in that it applies to any TGFT model (in particular quantum geometric ones, i.e. GFTs), in which the domain $\mathcal{D}$ of the mean field $\sigma(\mathcal{D})$ (and of the fundamental field $\varphi(\mathcal{D})$) is understood as the space of geometries of a single (spacelike) 3-simplex (or conjugate extrinsic geometry), plus additional data (like the scalar field of the example we are considering) that can be used to define physical reference frames. The general supporting fact is the following.
It can be shown \cite{Gielen:2014ila,Jercher:2021bie} that such domain $\mathcal{D}$ (space of geometries of a single 3-simplex, plus discrete matter values) is diffeomorphic to the space of metrics (or conjugate extrinsic curvatures) at a point in a 3d (spacelike) hypersurface, plus matter field values at the same point, which in turn is diffeomorphic to the minisuperspace of continuum homogeneous 3-geometries (or conjugate homogeneous extrinsic data), plus homogeneous matter fields.
This implies that the TGFT condensate wavefunction $\sigma(\mathcal{D})$ can be understood as a wavefunction on minisuperspace, as in quantum cosmology. 

\

However, the other general result is that this wavefunction satisfies non-linear dynamical equations, not the linear ones of quantum cosmology (i.e.the Wheeler-DeWitt equation restricted to wavefunctions on minisuperpsace). These are the quantum equations of motion derived from the quantum effective action or, in the simpler approximation we are focusing on, the classical equations of motion of the chosen TGFT model (the direct analogue of a Gross-Pitaevskii hydrodynamic equation for a quantum fluid):

\begin{equation}
\int [dg'] d\chi' \mathcal{K}\left( [g] , [g']; \chi, \chi' \right)\,\sigma(g',\chi')\, + \, \lambda \frac{\delta}{\delta\varphi^*} \mathcal{V}(\varphi , \varphi^*) _{|_{\varphi = \sigma}} = 0 \qquad , \label{HydroEqns}
\end{equation}

with analogous equation for the conjugate TGFT field.

They are hydrodynamic equations, thus their non-linearity is not surprising and in fact, mandatory, stemming from the presence of fundamental interactions among QG atoms. 
Further, these equations are in general also non-local on minisuperspace, as follows from the non-local nature of the same TGFT interactions.

\

We obtain, then, a non-linear and non-local extension of a quantum cosmological equation, for the condensate wavefunction, encoding an infinity of quantum gravity degrees of freedom in a coarse grained, collective manner. This is the more precise sense in which cosmology results from quantum gravity hydrodynamics and the universe is recognized as a quantum gravity condensate.

From the point of view of the emergence scheme of Fig.~\ref{fig:emergence}, the step from QG atoms to a continuum gravitational description (yet to be fleshed out in terms of observables), needed in any {\it level 1 emergence}, and leading to this cosmological setting, is their hydrodynamic approximation, then looked at in a specific (condensate) phase, and thus from a {\it level 2 perspective}. 

Let us stress however that the correspondence with quantum cosmology is limited to the kinematical aspects, since we have seen that the dynamics is necessarily non-linear in the condensate wavefunction. 
The distance from quantum cosmology is also important at the conceptual level. Despite the similarities, extending to the way geometric observables can be constructed, as we are going to see in the following, the fundamentally nonlinear dynamics prevents solutions to form a Hilbert space and thus any superposition principle and any quantum mechanical interpretation for the condensate wavefunction itself, as \lq quantum state of the universe\rq or the like. While the fundamental quantum gravity degrees of freedom governed by the TGFT model are treated quantum mechanically , with a HIlbert space given by the TGFT Fock space, the interpretation of the TGFT condensate wavefunction and mean field can only be of statistical and epistemic type, as appropriate for a coarse grained quantity like the density of a quantum fluid. 
This picture of cosmology as quantum gravity hydrodynamics raises then a number of conceptual issues also from the contrast with the traditional framework of quantum cosmology. We leave them for future analysis.

\subsection{Extracting effective spacetime/geometric dynamics}
The next task is to choose a specific (class of) GFT model(s) with a good interpretation from the point of view of discrete gravity (as it appears in its perturbative expansion, and then to recast the condensate hydrodynamic equations into equations for geometric observables and to give them a \lq temporal\rq form by implementing the relational strategy. In other words, looking at our general scheme for spacetime emergence in Fig. ~\ref{fig:emergence}, we need to move from the effective continuum formulation of full quantum gravity to some quantum version of General Relativity, looking at specific observables.

We consider quantum geometric models with action of the general form ~\ref{GFTaction}, and specifically Lorentzian EPRL-like models \cite{Perez:2012wv, Oriti:2016qtz} or the Lorentzian Barrett-Crane model \cite{Perez:2012wv,Jercher:2021bie}. We refer to the literature for the explicit form of the GFT action (or of the spin foam amplitudes from which the GFT action can be deduced) for these models. The first class of models adopts a formulation of Lorentzian quantum geometry based on a map from data from the Lorentz group to $SU(2)$ data \cite{Finocchiaro:2020xwr}, and it is closer to canonical LQG as well, while the second uses Lorentzian data only. We adopt a notation referring to $SU(2)$ data, for simplicity, but our results apply to both types of models.

We are interested in reconstructing homogeneous cosmological dynamics and we focus on isotropic configurations only. The simplest way to do so is to restrict the condensate wavefunction to isotropic data. This restriction \cite{Oriti:2016qtz} amounts to working with functions which depend on a single representation label, once expanded in irreps of the group, and on the scalar field values only: $\sigma^j(\chi)$, corresponding to the fact that one single metric degree of freedom, e.g. the universe volume (or the scale factor), is sufficient to determine the full geometric configuration of the universe. 

The second approximation we apply is to assume that GFT interactions are subdominant compared to the free GFT dynamics, to which we then restrict attention in the first place. This is consistent (in fact, required) by the focus on lattice gravity and spin foam dynamics to guide our physical interpretation of these GFT models (since both descriptions arise in the perturbative GFT regime), and with the use of simple coherent states of the GFT field operator (since strong interactions would generate strong quantum correlations among GFT quanta, which these states do not account for).

\

Next, we want to recast the hydrodynamic equations ~\ref{HydroEqns} for these models, restricted to isotropic wavefunctions, in a relational evolution form. To do so, we choose the scalar field degree of freedom $\chi$ as our relational clock and we consider condensate states that are \lq semiclassical enough\rq to admit such variable as a good clock.\footnote{Other constructions can be found in the literature, based on distributional relational operators (and generic coherent states \cite{Oriti:2016qtz,Pithis:2016cxg} or on a classical deparametrization of the GFT system with respect to the chosen scalar field clock \cite{Wilson-Ewing:2018mrp,Gielen:2021dlk,Gielen:2019kae,Gielen:2020fgi}. They give similar results. Another strategy for the extraction of effective relational dynamics \cite{Bojowald:2010xp}, not relying on specific choices of quantum states, can also be applied to the GFT system \cite{Gielen:2021vdd} and lead to a different choice of clock, making the comparison less straightforward.} We work with \lq coherent peaked states\rq \cite{Marchetti:2020umh} of the form:
\begin{equation} \label{CPS}
\sigma_\epsilon(j;\chi) \equiv \eta_\epsilon(j;\chi-\chi_0;\pi_0)\tilde{\sigma}(j;\chi)\qquad ,
\end{equation}
where $\eta$ is a function (e.g. Gaussian) peaked around the $\chi_0$ value of the clock variable $\chi$, with width given by $\epsilon \ll1$, and depending on a second parameter $\pi_0$ governing the fluctuations in the conjugate variable to $\chi$ (related to the momentum of the scalar field, whose fluctuations are small if $\pi_0^2\epsilon \gg 1$). Notice that this semiclassicality condition only refers to the clock values, while the geometric degrees of freedom can be highly quantum. The resulting relational temporal picture will then be approximate and effective only, since it results both from some coarse graining (it is only the collective observable corresponding to $\chi$ at the hydrodynamic level that will have the interpretation of (clock) time), and of neglecting some physical features of the system chosen as clock (the quantum properties of effective continuum scalar field).

The peaking function $\eta$ allows to approximate the remaining part of the condensate wavefunction $\tilde{\sigma}$ at $\chi_0$ and to neglect higher orders in the derivative expansion with respect to $\chi$ of the kinetic term in ~\ref{HydroEqns}. The resulting equation for $\tilde{\sigma}$ is:

\begin{equation} \label{EffEqn}
\tilde{\sigma}^{''}_j(\chi_0) \, -\, 2 i \tilde{\pi}_0 \tilde{\sigma}^{'}_j(\chi_0) \,-\,E_j^2 \,\tilde{\sigma}_j(\chi_0)\,=\,0 \qquad ,
\end{equation}
where the derivatives are with respect to $\chi_0$, $\tilde{\pi}_0 = \frac{\pi_0}{\epsilon \pi_0^2 - 1}$ and $E^2_j = \epsilon^{-1}\frac{2}{\epsilon \pi_0^2 - 1} + \frac{B_j}{A_j}$, with $A_j$ and $B_j$ being the coefficients of the 0th and 2nd order terms in the expansion of the kinetic term of the GFT action in terms of derivatives with respect to the scalar field values (thus, they are functions of the quantum geometric data $j$ and of the fundamental parameters defining the model itself).
This is the effective dynamical equation describing the evolution of the condensate function with respect to the clock time $\chi_0$. 

\

Now we have to recast it in spatiotemporal form, and extract from it a dynamical evolution equation for some relevant geometric observable. In the scheme of Fig.~\ref{fig:emergence}, it means going from a level 1 situation to a level 0 one, corresponding to (the cosmological sector of) quantum GR.

For doing so, we define the relational observables that we expect to be relevant for describing homogeneous cosmological evolution \cite{Marchetti:2020umh}. These are expectation values of fundamental GFT operators (acting on the GFT Fock space), evaluated on the coherent peaked states ~\ref{CPS}, and thus well approximated by the value the condensate wavefunction takes at $\chi=\chi_0$. The geometric interpretation is guided by the discrete gravity picture of quantum states and perturbative GFT amplitudes.
We have the occupation number: 

\begin{equation}
N(\chi_0) \equiv \langle \widehat{N}\rangle_{\sigma;\chi_0,\pi_0} = \sum_j \rho_j^2(\chi_0) \qquad ,
\end{equation}

the universe volume (constructed from the matrix elements of the 1st quantized volume operator for GFT quanta, i.e. quantized tetrahedra, with eigenvalues $V_j$, convoluted with field operators\footnote{In other words, the operator adds the individual volume contributions from the GFT quanta populating the state.}):
\begin{equation}
V(\chi_0) \equiv \langle \widehat{V}\rangle_{\sigma;\chi_0,\pi_0} = \sum_j V_j \rho_j^2(\chi_0) \qquad ,
\end{equation}

the clock (scalar field) value:

\begin{equation}
\frac{\langle \widehat{\chi}\rangle_{\sigma;\chi_0,\pi_0}}{N(\chi_0)} \simeq \chi_0 \qquad ,
\end{equation}

and the scalar field momentum:
\begin{equation}
\langle \widehat{\Pi}\rangle_{\sigma;\chi_0,\pi_0} \simeq \pi_0\left( \frac{1}{\epsilon \pi_0^2 - 1} + 1\right) N(\chi_0) + \sum_j Q_j \qquad ,
\end{equation}
where $Q_j$ are conserved quantities, and we have adopted the decomposition of the condensate wavefunction in terms of standard hydrodynamic variables: the density of the fluid and the phase (from which, in standard hydrodynamics, one defines the fluid velocity): $\tilde{\sigma}_j(\chi) = \rho_j(\chi) e^{i\theta_j(\chi)}$. All these observables are clearly geometric and spatiotemporal only in an approximate, coarse-grained and collective sense. 

\

Using the hydrodynamic equation ~\ref{EffEqn}, we obtain the equations governing the relational evolution of the universe volume \cite{Marchetti:2020umh,Oriti:2016qtz}:
\begin{equation} \label{QuantumFriedmann}
\left( \frac{V'}{3V}\right)^2 \simeq \left(\frac{2 \sum_j V_j \rho_j sgn(\rho^{'}_j) \sqrt{\mathcal{E}_j - Q^2_j/\rho^2_j + \mu^2_j \rho^2_j}}{3\sum_j V_j \rho^2_j} \right)^2 \hspace{1cm} \frac{V^{''}}{V} \simeq \frac{2 \sum_j V_j \left[ \mathcal{E}_j + 2 \mu^2_j \rho_j^2\right]}{\sum_j V_j \rho_j^2} \qquad ,
\end{equation}
where we have defined $\mu^2_j = E^2_j - \tilde{\pi}_0^2$ and $\mathcal{E}_j$ is another conserved quantity. 

These are the {\it generalised (quantum-corrected) Friedmann equations in relational time} (given by the scalar field value $\chi_0$) that our quantum gravity model gives for the emergent spacetime in the homogeneous case (as captured by the volume observable, in the presence of a real, massless, free scalar field as the only matter content of the universe). 

We are now at level 0 of spacetime emergence, in our scheme (Fig. ~\ref{fig:emergence}), that of quantum GR. Our geometric, spacetime notions (only the volume, in this simple case) are well defined and under control, but satisfy quantum-corrected equations, are subject to (possibly) strong quantum fluctuations, etc.

\

To recover standard notions of space and time (just the usual notion volume, here), we should recover the classical equations it satisfies in GR, and check that the quantum fluctuations become negligible, in the same limit.
The classical regime corresponds to the one in which the Hubble rate is small compared to the inverse Planck time, i.e. small curvature; while technically more subtle, it is reached for large universe volumes. In this regime, in which the QG fluid density is large compared to the other terms appearing in the equations, we get the approximate dynamical relations:
\begin{equation} \label{AlmostFriedmann}
\left( \frac{V'}{3V}\right)^2 \simeq \left(\frac{2 \sum_j \mu_j V_j \rho_j^2 sgn(\rho^{'}_j}{3\sum_j V_j \rho^2_j} \right)^2 \hspace{1cm} \frac{V^{''}}{V} \simeq \frac{4 \sum_j V_j \left[ \mu^2_j \rho_j^2\right]}{\sum_j V_j \rho_j^2} \qquad .
\end{equation}
This means that for any GFT model (in the class we considered) in which the right-hand-side of both equations becomes approximately a constant $g$, we recover the usual Friedmann equations in relational time:
\begin{equation}\label{Friedmann}
\left( \frac{V^{'}}{V}\right)^2 = \frac{V^{''}}{V} = 12 \pi \, G \qquad
\end{equation} 
by identifying the effective constant $g$ with Newton's constant (up to a numerical factor). This is the case, for example, if there is a single dominant mode $j$, in which case $\mu_j^2 = 3 \pi G$ or if $\mu_j^2 = 3 \pi G \; \forall j$. This we have the usual solutions corresponding to an expanding universe of the standard model of cosmology. 

In fact, also all the conditions that need to be verified on the quantum fluctuations of the relational clock (that has to remain semiclassical to have a good notion of temporal evolution) and of the universe volume (that have to become negligible) can be checked explicitly \cite{Marchetti:2020qsq}. In this regime, thus, we recover standard spacetime, and its emergence from full quantum gravity is complete, having realized that the proper notion of time (and space) should be defined in terms of physical frames (and not of manifold structures, which can at most play an auxiliary role). We are at level -1, in our schematic classification, the realm of classical continuum GR. 
 
One thing we learn, looking at GR now from this emergent spacetime perspective, is that gravitational parameters, like Newton's constant, are in fact functions of the microscopic, fundamental quantum gravity parameters, those characterizing the quantum gravity atoms and their quantum dynamics. This is because $\mu_j^2$, which becomes identified with Netwon's constant in the large-volume, late-time regime of our emergent cosmological evolution, is a function of these parameters. This is nothing surprising, since exactly the same happens in standard hydrodynamics, for example, in its relation to the atomic or molecular theory underlying it.

\subsection{Cosmological singularity and geometrogenesis}
This is of course not the only thing we learn, or we expect to learn, from a full quantum gravity account of the emergence of space and time from fundamentally non-spatiotemporal entities. Let us discuss a few more results obtained in this context, and the lessons we learn from them.

\

First, we want our theory of quantum gravity to tell us what happens to the cosmological singularity, that signals the breakdown of classical GR (or at least, whose physics cannot be accounted for by classical GR).
Looking again at our quantum cosmological dynamics ~\ref{QuantumFriedmann}, we realize an important fact. Under the assumption that the \lq good-clock conditions\rq (small fluctuations in the clock value and in the scalar field momentum) remain satisfied, If at least one conserved quantity $Q_j$ or $\mathcal{E}_j$ is non-vanishing, there exist one value $j$ such that the fluid density $\rho_j(\chi_0)$ remains different from zero at all times $\chi_0$. This implies that our expanding universe can be followed toward earlier times to find that its volume remains always positive and with a single turning point. That is, we find a quantum bounce instead of the big bang (thus solving the classical cosmological singularity).\footnote{This behaviour can be recast in the language of an effective mimetic gravity theory \cite{deCesare:2018cts}.} 
Moreover, one can compute what happens to quantum fluctuations in the early universe \cite{Marchetti:2020qsq}, to find that they remain small at least for a specific but rather large class of solutions to the quantum dynamics.

\

Both results become particularly transparent for the simplest quantum gravity condensates corresponding to condensate wavefunctions which are non-vanishing only for a single mode $j_0$. For them, identifying $\mu_{j_0}^2 = 3 \pi G$, and with all $Q_j$ vanishing, we find the cosmological equation:

\begin{equation} \label{SimpleCondensate}
\left[ \frac{V^{'}}{3V}\right]^2 = \frac{4 \pi G}{3} + \frac{4 V_{j_0} \mathcal{E}_{j_0}}{9 V} \qquad V_{j_0} \approx V_{Planck} j_0^{3/2}
\end{equation}
such that all solutions corresponding to $\mathcal{E}_{j_0} < 0$ describe a quantum bounce at $V_{min} = V_{j_0} N_{min} = \frac{V_{j_0} |\mathcal{E}_{j_0}|}{6 \pi G} $. The relative volume fluctuations, on the other hand, behave as $\frac{\Delta V}{V}(\chi_0) \approx \frac{1}{N(\chi_0)}$, so they remain small provided the (average) occupation number does not become of order one.

\

Before discussing in more detail the fate of the cosmological singularity and the beginning of our universe in an emergent spacetime scenario, let us mention a couple of recent research directions showing that the consequences of such scenario do not have to be confined to the extreme conditions of the very early universe. 

Cosmological perturbations, thus inhomogeneities in both geometry and matter, are important for making stronger contact with cosmological observations and because it is at the level of their field theory description that the physical aspects of spacetime dynamics can truly be probed, in their quantum gravity origin. Moreover, it is only when reproducing local spacetime features that any quantum gravity model can be said to provide an example of an emergent space and time. Cosmological perturbations in a GFT cosmology context are therefore receiving increasing attention \cite{Gielen:2017eco,Gielen:2018xph,Marchetti:2021gcv}. 

From the point of view of spacetime emergence, two points need to be noted, of this recent work. First, the notion of spacetime point, and thus local physics, needs to be defined relationally, in terms of appropriate physical degrees of freedom, the easiest choice in the GFT context being four massless real scalar fields, used as rods and clock, extending the relational framework developed to have a notion of temporal evolution in the homogeneous case. Second, cosmological perturbations can be identified with (relationally local) excitations over a homogeneous GFT condensate, in direct analogy with quasi-particles in a quantum fluid, and can be studied (at first) in the same mean field approximation \cite{Marchetti:2021gcv}. 

The lesson is that the framework of effective (quantum) field theory on a given (quantum) background can be reproduced from (perturbative) QG hydrodynamics.\footnote{In fact, once appropriate matter degrees of freedom are coupled to quantum geometry in the fundamental QG model, the hydrodynamic approximation encodes both the data needed to reconstruct the metric at a point and the ones allowing to define points in a diffeo-invariant manner. Therefore, there is no obstacle to attempt a reconstruction of all gravitational dynamics and full GR from it, without confining ourselves to a perturbative regime. Whether a mean field approximation is enough to recover the correct physics is, of course, a different story.}  

Late-time cosmology is also a testbed for new physics, since the observed acceleration of cosmological expansion remains puzzling from many reasons \cite{Brax:2017idh,Burgess:2013ara}. Being a large-scale phenomenon, this is difficult to understand from the point of view of microphysics and not usually understood as originating from quantum gravity, although a more fundamental theory is certainly called to contribute to its understanding. 

From an emergent spacetime perspective, however, the whole spacetime dynamics at both small and large scales is in fact of direct quantum gravity origin and we are encouraged to think outside the usual effective field theory mindset\footnote{The very notion of scales and separation of scales can be said to be of very dubious meaning in a background independent context, and even more so in an emergent spacetime scenario.} (see also supporting results in the analogue gravity context \cite{Finazzi:2011zw}). In the GFT cosmology context, recent work shows that the macroscopic effect of the fundamental interactions among QG atoms (that we neglected in our analysis above) can be relevant and producing interesting consequences \cite{Pithis:2016wzf,Pithis:2016cxg,Gielen:2019kae}. 

In particular, and quite strikingly, they can produce quite naturally an accelerated dynamics at late times of a phantom dark energy type \cite{Oriti:2021rvm}. It is too early to consider this result as a full, compelling explanation of dark energy, and more work is needed to do so. 
However, it does already constitute a proof-of-principle from which we can draw one main lesson: in an emergent spacetime scenario, effective field theory intuition is bound to fail and large scale physics can be of direct quantum gravity origin.

\

Let us now go back to the fate of the cosmological singularity in GFT cosmology (and more generally, in an emergent spacetime scenario). What happens to the cosmological singularity in quantum gravity, in light of GFT cosmology?

We have seen that the volume evolution governed by the equations ~\ref{QuantumFriedmann} (and, in the simplest case, ~\ref{SimpleCondensate}), contains quantum gravity corrections akin to a sort of \lq quantum pressure\rq preventing it from reaching vanishing configurations, so that the classical cosmological singularity is replaced by a quantum bounce.\footnote{It must be noted that this quantum pressure, while very similar in its effects to the one found in loop quantum cosmology, can be traced back to a never-vanishing number density, in this GFT hydrodynamics context, rather than to the discreteness of volume spectrum or absence of zero eigenvalues from it (in fact, it is present also for GFT models where the volume spectrum is continuous \cite{Jercher:2021bie}).} More precisely, we found that the big bang singularity is replaced by a big bounce scenario in a mean field restriction of the hydrodynamic approximation to the full quantum dynamics, within a condensate phase.

The mean field approximation should obviously be improved, but it could well be that the bouncing scenario is stable under such improvements. If that turns out to be the case, then we could really say that quantum gravity (better, GFT cosmology) predicts a cosmic quantum bounce. 

\

Is that it? Not quite. This would be the conclusion only if the hydrodynamic approximation still holds and remains reliable at the would-be bounce, and if the whole quantum gravity system stays within the condensate phase. 
But even before considering these extreme possibilities, even within the hydrodynamic approximation we need a well-behaved clock to speak of evolution towards and then across the bounce. If the clock becomes subject to too strong quantum fluctuations, for example, we could have no reliable notion of evolution, and thus we could only follow the dynamics up to the point where the very notion of time and evolution ceases to make sense. 

Suppose however that this does not happen. It could still be the case that fluctuations become too strong, at the would-be bounce, to invalidate the hydrodynamic approximation within which we have been able to extract a geometric, spatiotemporal description for our system of QG atoms.
Then the best we could say would be, again, that the universe history can be followed backward from present day up to a point in which space and time simply disappear, we do not have a reliable spatiotemporal or geometric description, and we have to resort to the more fundamental, non-spatiotemporal description to obtain needed input for understanding the new physics of the origin of the universe. What replaces the big bang singularity, then, is the disappearance of spacetime (when read \lq backward in time\rq) or its emergence (when read \lq forward in time\rq).

It could still be the case that the fundamental microscopic quantum gravity theory gives us a complement to the hydrodynamic description that can still be somehow translated in those terms. One could try, for example, to add terms coming from kinetic theory to standard hydrodynamics, in order to improve the physical description, while maintaining the same intuition about the physical entities one is dealing with. Then we would know that the spatiotemporal description is not entirely valid in that regime, and that there is no bounce, actually, but we could still describe the effects of the non-spatiotemporal dynamics in the corresponding regime in spatiotemporal terms.

\

Not so if the quantum fluctuations affect so drastically the fundamental dynamics and the behaviour of (geometric) observables to drive the system out of the geometric, spatiotemporal phase (i.e. the condensate phase in GFT cosmology) to the phase transition separating it from a non-geometric one.
Then we would face an even more radical disappearance of spacetime, and the non-spatiotemporal description in terms of the fundamental QG atoms would be necessary to capture the relevant physics, being also not translatable into spacetime language.
The phase transition between non-geometric and geometric phases of the quantum gravity system would be then what truly replaces the big bang of continuum and classical GR, in this quantum gravity scenario. Geometrogenesis replaces the big bang in quantum gravity.

From the point of view of our scheme ~\ref{fig:emergence} we are then led to exploring the details of the continuum phase diagram and of the phase transitions it contains. From the conceptual point of view, we are then confronting a level 3 emergence of spacetime, in which the philosophical issues concerning the geometrogenesis phase transition should be tackled, alongside the physical ones.
The two sets of issues are intertwined, and intertwined as well with the physics of cosmological evolution and the early universe. This means that, for example, in the TGFT context,  the analysis of the RG flow of (ideally) fully quantum geometric models should be performed in parallel with the analysis fo their emergent cosmological dynamics, with the two research lines informing each other. 

\

We can illustrate this relation in a simple case, albeit in rather sketchy manner.
This example shows how one could give a temporal characterization to a geometrogenesis phase transition, at least in the sense of localizing it in time with respect to the geometric phase, in particular making concrete the identification with the regime corresponding to the big bang singularity in the classical continuum case, and to the quantum bounce in the GFT hydrodynamic approximation.

Consider the cosmological evolution for the simplest single-mode condensates of Eq.~\ref{SimpleCondensate}. Given the expression for the volume at the bounce and of the quantum fluctuations, it is clear that these quantum fluctuations are indeed maximal at the bounce and given by:
\begin{equation}
\left[ \frac{\Delta V}{V}(\chi_{crit})\right]_{max} \approx \frac{1}{N_{min}(\chi_{crit})} \simeq \frac{G}{\mathcal{E}_{j_0}(G)}
\end{equation}
where we have indicated that the conserved quantity $\mathcal{E}$ is in fact a function of the parameters of the model and thus of the effective Newton constant $G$. 

The point is that we expect the hydrodynamic approximation to break down when the average occupation number becomes smaller than one, and indeed this is also when the relative fluctuations in the universe volume become larger than one, i.e. for $\frac{G}{\mathcal{E}_{j_0}(G)} \geq 1$.
In turn we should recall that the effective Newton constant is, really, a function of the fundamental couplings of the underlying GFT model, i.e. $G = G(\{ \lambda_i\})$.

One way in which a geometrogenesis phase transition can then be associated to the physical origin fo the universe evolution is if the critical regime fo the GFT couplings corresponding to the phase transition are shown to give $\frac{G}{\mathcal{E}_{j_0}(G)} \geq 1$, signalling the breakdown of the cosmological dynamics extracted in the hydrodynamic approximation. This should also happen in the regime that, in the same hydrodynamic approximation, would correspond to the quantum bounce.

This last identification requires, on the other hand, a clear relation between the RG flow parameter used in the RG analysis of the GFT model and the geometric data used to given spatiotemporal meaning to the GFT hydrodynamic equations. 
In general, for GFT models based on both quantum geometric and scalar matter data (i.e. like the ones we focused on here) the RG scale would be a combination of both. To simplify the picture, let us assume that the RG scale, i.e. the \lq IR cut-ff\rq in a full functional RG analysis \cite{Carrozza:2016vsq,Baloitcha:2020idd, Finocchiaro:2020fhl,Pithis:2020kio} or a mean field one \cite{Marchetti:2021fix} is just given by a spin label $k$. 

The quantum bounce found in the hydrodynamic approximation has to take place, given observational constraints, not too distant from Planckian volumes, thus the minimal spin label giving $V_{min} = V_{j_0} N_{min}$ should be close to lowest end of the volume spectrum (and the lowest possible, to try to avoid $N_{min}$ having to be necessarily too small).\footnote{Obviously, this is very heuristic and we are neglecting a number of aspects of a more realistic treatment, not last the fact that we are using the effective dynamics obtained considering condensate wavefunctions with only one spin excitation $j_0$, while a non-trivial RG flow requires of course a non-trivial field dependence on the same modes $j$.} 

This implies that the hydrodynamic equation would have to be used close to the \lq IR end\rq of the RG flow of the same GFT model, where the GFT couplings (at least some fo them) will reach their critical values. The would-be quantum bounce, then, will happen exactly close to the critical regime of the underlying quantum gravity model.
If it happens that, going to such critical values of the couplings, we have

\begin{equation}
\frac{G(\lambda)}{\mathcal{E}_{j_0}(G(\lambda))} \longrightarrow \frac{G(\lambda_{crit})}{\mathcal{E}_{j_0}(G(\lambda_{crit}))} \geq 1
\end{equation} 
then the fluctuations grow too much, preventing the quantum bounce to happen, or, better, making it physically irrelevant, since the hydrodynamic approximation in which it is formulated would not be reliable. The underlying physics would then be instead characterized by the properties of the GFT phase transition, the geometrogenesis.

\

Let us recapitulate the logic of this tentative scenario. The geometrogenesis phase transition is a transition between two phases (not a phase in itself), a geometric phase and a non-geometric one, by definition. The question I am posing in the text is whether this transition can be somehow localised in time and I answer that this can in principle be done but only from the standpoint of the geometric phase we live in, as one could intuitively expect, since there is no notion of time that could be applied to the whole set of continuum phases, thus encompassing both geometric and non-geometric phases, or to the level of description in terms of TGFT quanta. For example, it could be localised as having happened in our past and identified with the very early stages of the evolution of the universe, if one finds that the fundamental theory locates at that time (within an effective cosmological dynamics) the regime of very strong fluctuations in geometric observables that characterizes the phase transition. In this case, we have to conclude that a geometrogenesis scenario is more appropriate than a bouncing scenario, from the perspective of the fundamental quantum gravity theory, as the appropriate account of the very early universe and as a replacement of the classical big bang singularity.

\

All the above reasoning is tentative as much as it is sketchy, and obviously much remains to be done to put it on solid grounds, both concerning the RG analysis of TGFT models and their effective cosmological (more generally, gravitational) continuum dynamics.
It shows however that even issues associated to {\it level 3 spacetime emergence}, exotic as they may be, can be in principle tackled in very concrete terms, within the TGFT formalism. 

\section{Conclusions: many more questions, but one concrete framework to investigate them}
What we have just concluded about the geometrogenesis scenario is in fact the main message of this contribution also for spacetime emergence in general.

There are many open issues, both technical and physical, concerning the TGFT cosmology framework. We still need to go convincingly beyond homogeneity and isotropy, and towards extracting the full continuum gravitational dynamics from the fundamental quantum gravity theory (based on a number of approximations and assumptions, for sure). We need to investigate in more detail several aspects of the cosmological evolution we extracted so far (both in the very early universe and at late times), improving the approximations on which it is based (mean field hydrodynamics), and to enrich it with more physical ingredients (e.g. realistic matter content), to see if quantum gravity can truly solve current cosmological puzzles. More associated physical issues could be named, for example the precise relation between the phenomena described in different physical (relational) frames.
The important point is that all of them can be tackled in detail (at least in principle) and very concretely in this template for spacetime emergence in quantum gravity.

The same is true for the many open issues at the philosophical level, concerning foundations of spacetime and quantum gravity, and of physics more generally. The framework of TGFT cosmology offers a concrete testbed for philosophical analyses concerning the nature of space and time and their emergence, of course, i.e. the focus of this contribution, at both metaphysical and epistemological levels. Indeed, a partial list of interesting research directions include: 
the development of a new ontology that is not grounded on space and time notions (a tantalizing as much as a daunting challenge for metaphysics); the analysis of the role of observers and agents in emergent spacetime scenarios and, more generally, in quantum gravity, where the absence of preferred or classical notions of time evolution and causal structure complicates the debate on the nature of probabilities, and where the remote character of the relevant physical phenomena makes any operational approach to physical theories questionable, while at the same time putting under strain any naive picture of physical theories as purely representational and objective (since only very indirectly grounded on empirical data); the analysis of the possible perspectival nature of cosmological evolution (and, for example, of a thermodynamic time arrow for such evolution), since this is can only be understood in relational terms and thus referring to a specific internal (to the universe) physical frame, and in a context in which the elements that are left invariant by the switch to another internal, physical frame are not fully under control; more generally, the issue of  how all the above elements should affect our understanding of laws of nature, since it is not obvious that any of the traditional accounts (Humeanism, primitivism, universals-based, best-systems, etc) can be applied in absence of space and time as background conceptual structures, and what would be left, more generally, of any traditional form of realism, in the same context.

We hope that this contribution, if it has not provided answers to those issues, has at least made clear that these answers can be found, in principle, within the quantum gravity context we presented.   
 
\section*{Acknowledgements}
We acknowledge funding from DFG research grants OR432/3-1 and OR432/4-1. We also express our gratitude to the editor of this volume for the invitation to contribute to it and for the remarkable patience demonstrated in dealing with us. 
 
\bibliographystyle{jhep}
\bibliography{Warsaw-Bib}

\end{document}